\pdfoutput=1

\documentclass[10pt,twocolumn,prd,aps,amssymb,amsmath,tightenlines,showpacs]{revtex4}

\usepackage{graphicx}

\newcommand{\cP}{\ensuremath{\mathcal{P}}}
\newcommand{\cT}{\ensuremath{\mathcal{T}}}

\newcommand{\e}{\ensuremath{\eta}}
\newcommand{\eb}{\ensuremath{\bar{\eta}}}

\begin{document}

\title{$\cP\cT$-Symmetric Representations of Fermionic Algebras}

\author{Carl M. Bender$^a$}\email{cmb@wustl.edu}
\author{S. P. Klevansky$^b$}\email{spk@physik.uni-heidelberg.de}
\affiliation{$^a$Physics Department, Washington University, St. Louis, MO 63130,
USA}
\affiliation{$^b$Institut f\"ur Theoretische Physik, Universit\"at Heidelberg,
Philosophenweg 19, 69120 Heidelberg, Germany}

\date{\today}

\begin{abstract}
A recent paper by Jones-Smith and Mathur extends $\cP\cT$-symmetric quantum
mechanics from bosonic systems (systems for which $\cT^2={\bf 1}$) to fermionic
systems (systems for which $\cT^2=-{\bf 1}$). The current paper shows how the
formalism developed by Jones-Smith and Mathur can be used to construct $\cP
\cT$-symmetric matrix representations for operator algebras of the form $\e^2=
0$, $\eb^2=0$, $\e\eb+\eb\e=\alpha{\bf 1}$, where $\eb=\e^{\cP\cT}=\cP\cT\e
\cT^{-1}\cP^{-1}$. It is easy to construct matrix representations for the
Grassmann algebra ($\alpha=0$). However, one can only construct matrix
representations for the fermionic operator algebra ($\alpha\neq0$) if $\alpha=
-1$; a matrix representation does not exist for the conventional value $\alpha=
1$.
\end{abstract}


\maketitle

\section{Introduction}
\label{s1}
A recent paper \cite{X1} shows how to generalize $\cP\cT$ quantum mechanics from
the Heisenberg algebra $[q,p]=i$ to other kinds of algebras, such as E2. (The E2
algebra is characterized by a set of three commutation relations: $[u,J]=iv$,
$[v,J]=-iu$, $[u,v]=0$.) The algebras considered in Ref.~\cite{X1} are bosonic
in character because they are expressed in terms of commutation relations.
However, Jones-Smith and Mathur have shown how to describe $\cP\cT$-symmetric
quantum theories in a fermionic setting \cite{X2}. Thus, in the current paper we
apply the formalism developed in Ref.~\cite{X2} to examine the representations
of algebras expressed in terms of anticommutation relations.

We consider here two standard algebras: the operator algebra of fermions, which
consists of two nilpotent elements, $\e$ and $\eb$, whose anticommutator is
unity,
\begin{equation}
\e^2=0,\quad\eb^2=0,\quad\e\eb+\eb\e={\bf 1},
\label{e1}
\end{equation}
and the {\it Grassmann} algebra, which again consists of two nilpotent
elements, $\e$ and $\eb$, whose anticommutator vanishes,
\begin{equation}
\e^2=0,\quad\eb^2=0,\quad\e\eb+\eb\e=0.
\label{e2}
\end{equation}
The requirement that $\e$ and $\eb$ be nilpotent is imposed to incorporate 
fermionic statistics. Our objective is to find matrix representations of
(\ref{e1}) and (\ref{e2}) in the context of fermionic $\cP\cT$-symmetric quantum
mechanics; that is, under the assumption that $\eb$ is the $\cP\cT$ reflection
of $\e$:
\begin{equation}
\eb=\cP\cT\e\cT^{-1}\cP^{-1}.
\label{e3}
\end{equation}

In Sec.~\ref{s2} we investigate the two-dimensional matrix representations of
(\ref{e1}) and (\ref{e2}) using nothing more than the representation of $\cP$
and $\cT$ introduced in Ref.~\cite{X3} in which the square of the $\cT$
operator is unity: $\cT^2={\bf 1}$. Then in Sec.~\ref{s3}, we apply the more
elaborate formalism introduced recently in Ref.~\cite{X2} in which is it argued
that $\cT^2=-{\bf 1}$ (and not ${\bf 1}$) for fermions. We show that if we
replace the condition $\cT^2={\bf 1}$ by $\cT^2=-{\bf 1}$, it is still possible
to find matrix representations of the Grassmann algebra (\ref{e2}). However, the
surprise is that it is not possible to find matrix representations of the
fermion operator algebra in (\ref{e1}); one can only find matrix representations
of the $\cP\cT$ version of the fermionic operator algebra
\begin{equation}
\e^2=0,\quad\eb^2=0,\quad\e\eb+\eb\e=-{\bf 1}.
\label{e4}
\end{equation}

\section{Two-dimensional representations of $\e$ and $\eb$}
\label{s2}

\begin{figure}[t!]
\begin{center}
\includegraphics[scale=0.46]{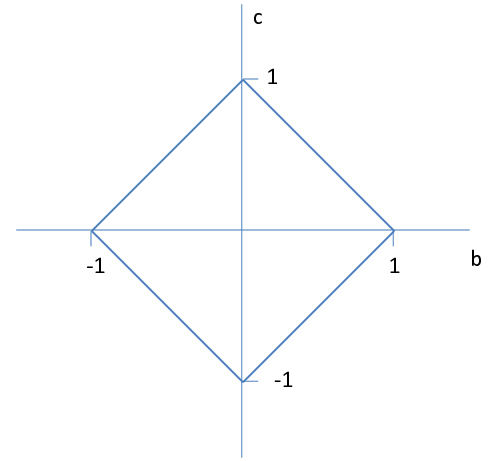}
\end{center}
\caption{Plot of the real parameters $b$ and $c$ for the representation of $\e$
in (\ref{e5}) for the case in which $\eb$ is the conventional Hermitian
conjugate of $\e$; $\e$ and $\eb$ satisfy the fermionic algebra $\e\eb+\eb\e=
{\bf 1}$. The diamond-shaped locus of points is described by the equation $|b|+
|c|=1$.}
\label{F1}
\end{figure}

For purposes of comparison, we begin by considering a two-dimensional matrix
representation in which we assume that $\eb$ is given by the conventional
Hermitian adjoint $\eb=\e^\dag$. The most general complex matrix $\e$ whose
square vanishes has vanishing trace and determinant,
\begin{equation}
\e=\left(\begin{array}{cc} a&b\cr c&-a\cr\end{array}\right),
\label{e5}
\end{equation}
where $b$ and $c$ are arbitrary complex numbers and $a$ is fixed by the
determinant condition $a^2+bc=0$. Then, $\eb$ is given by
\begin{equation}
\eb=\left(\begin{array}{cc} a^*&c^*\cr b^*&-a^*\cr\end{array}\right)
\label{e6}
\end{equation}
and the nilpotency condition $\eb^2=0$ is automatically satisfied. The fermionic
algebra condition $\e\eb+\eb\e={\bf 1}$ now reduces to
\begin{equation}
|b|+|c|=1.
\label{e7}
\end{equation}
Thus, if $b$ and $c$ are real, they are constrained to a unit diamond, as shown
in Fig.~\ref{F1}. More generally, if $b=ue^{i\alpha}$ and $c=ve^{i\beta}$ are
complex, then $\alpha$ and $\beta$ are arbitrary and $u\geq0$ and $v\geq0$ lie
on the line segment $u+v=1$ in the positive quadrant of the $(u,v)$ plane.

For the case of the Grassmann algebra, $b$ and $c$ satisfy the constraint
\begin{equation}
|b|+|c|=0
\label{e8}
\end{equation}
instead of (\ref{e7}). The unique solution to (\ref{e8}) is $b=c=0$. Thus, in
this case there is no nontrivial Grassmann representation for $\e$ and $\eb$.

Now let us turn to the case of a $\cP\cT$-symmetric fermionic algebra. What
happens if we apply to fermions the naive representations of parity reflection
$\cP$ and time reversal $\cT$ that were used earlier in Ref.~\cite{X3}? We
represent a parity reflection as a real symmetric matrix whose square is unity,
which for two-dimensional matrices is 
\begin{equation}
\cP=\left(\begin{array}{cc} 0&1\cr 1&0\cr\end{array}\right)
\label{e9}
\end{equation}
and we represent $\cT$ as complex conjugation. With these choices, $\cP^2=1$,
$\cT^2=1$, and $[\cP,\cT]=0$.

\begin{figure}[t!]
\begin{center}
\includegraphics[scale=0.46]{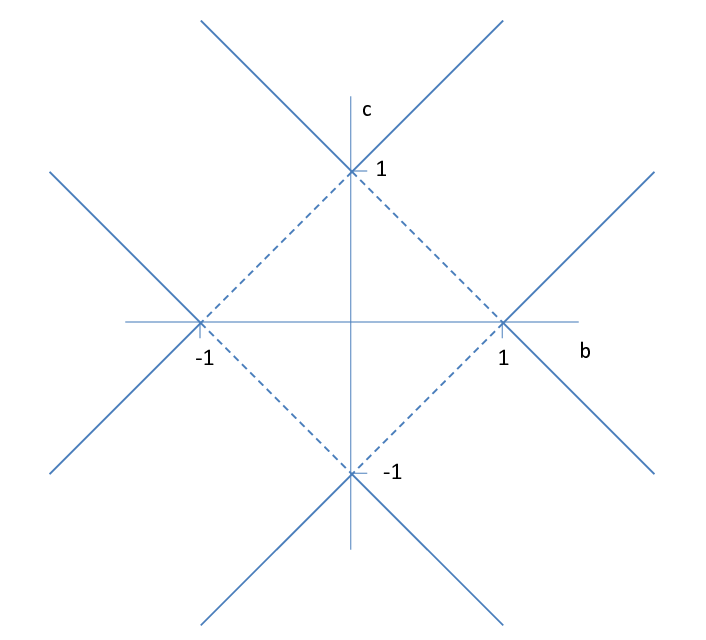}
\end{center}
\caption{Plot of the real parameters $b$ and $c$ for the representation of $\e$
in (\ref{e5}) for the case in which $\eb$ is the $\cP\cT$ conjugate of $\e$,
where $\cP$ is given in (\ref{e9}) and $\cT$ is complex conjugation, $\e$ and
$\eb$ satisfy a fermionic algebra. The locus of points is described by the
equation $(|b|-|c|)^2=1$ and is thus an unbounded region, in contrast with the
bounded region shown in Fig.~\ref{F1}.}
\label{F2}
\end{figure}

Here, if $\e$ is as given in (\ref{e5}), then from (\ref{e3}) $\eb$ is given by
\begin{equation}
\eb=\cP\cT\e\cP\cT=\left(\begin{array}{cc} -a^*&c^*\cr b^*&a^*\cr\end{array}
\right).
\label{e10}
\end{equation}
Once again, the condition $\eb^2=0$ is automatically satisfied. Now, requiring
that $\e$ and $\eb$ obey the fermionic algebra $\e\eb+\eb\e=1$ leads to the
condition
\begin{equation}
(|b|-|c|)^2=1.
\label{e11}
\end{equation}
Thus, if $b$ and $c$ are real, they lie on the lines shown in Fig.~\ref{F2},
which is the unbounded extended complement of the diamond shown in
Fig.~\ref{F2}. More generally, if $b=ue^{i\alpha}$ and $c=ve^{i\beta}$ are
complex, then $\alpha$ and $\beta$ are arbitrary, and $u\geq0$ and $v\geq0$ lie
on two infinite lines $u-v=\pm1$ in the positive quadrant of the $(u,v)$ plane.

If $\e$ and $\eb$ are required to satisfy a Grassmann algebra, then if $b$ and
$c$ are real, they must satisfy the equation
\begin{equation}
|b|-|c|=0.
\label{e12}
\end{equation}
Thus, unlike the Hermitian case, there is a nontrivial set of solutions.

It is interesting that when $\eb$ is the $\cP\cT$ conjugate of $\e$, there is an
unbounded range of parameters and that when $\eb$ is the Hermitian conjugate of
$\e$, the range of parameters is bounded. This result is strongly analogous to
what was found in the study of the $\cP\cT$-symmetric quantum brachistochrone
compared with the conventional Hermitian quantum brachistochrone \cite{X4}. The
matrix elements of the Hamiltonian that describes the $\cP\cT$ brachistochrone
are unbounded (even though the eigenvalues are fixed), while the matrix elements
of the Hamiltonian for the Hermitian quantum brachistochrone are bounded. Thus,
$\cP\cT$-symmetric quantum mechanics is hyperbolic (unbounded) in character,
while conventional Hermitian quantum mechanics is elliptic (bounded) in
character.

\section{Application of the Formalism of Jones-Smith and Mathur}
\label{s3}

For a correct quantum-mechanical description of fermions, the time-reflection
operator $\cT$ must be chosen such that its square is $-{\bf 1}$ instead of
${\bf 1}$ \cite{X5}. In a recent paper by Jones-Smith and Mathur, this fact is
used to construct suitable matrix representations of the $\cT$ and $\cP$
operators \cite{X2}. In the following subsection we briefly recapitulate their
results. 

\subsection{Brief summary of the essential results of Jones-Smith and Mathur}
\label{ss31}

In Ref.~\cite{X2} it is shown how to construct matrix representations of
dimension $4n$ $(n=1,\,2,\,3,\,\ldots)$ of the $\cT$ and $\cP$ operators. The
effect of a time operator acting on a state $\psi$, which is a $4m$-dimensional
vector, is to take the complex conjugate of $\psi$ and to multiply the result by
a real matrix $t$:
\begin{equation}
\cT\psi=t\psi^*.
\label{e13}
\end{equation}
The general form for the $t$ matrix consists of $2n$ copies of the $2\times2$
matrix 
$$\left(\begin{array}{cc} 0&1\cr -1&0\cr\end{array} \right)$$
on the main diagonal and zero entries elsewhere. For the simplest ($n=1$) case
$t$ is the $4\times4$ matrix
\begin{equation}
t=\left(\begin{array}{cccc}
0&1&0&0\cr -1&0&0&0\cr 0&0&0&1\cr 0&0&-1&0\cr\end{array}\right).
\label{e14}
\end{equation}
The effect of a parity operator acting on a state $\psi$ is to multiply $\psi$
by a real matrix $p$:
\begin{equation}
\cP\psi=p\psi.
\label{e15}
\end{equation}
The general form for the $p$ matrix is the diagonal matrix whose first $2n$
diagonal elements are $1$ and whose next $2n$ diagonal elements are $-1$. For
the simplest ($n=1$) case $p$ is the $4\times4$ matrix
\begin{equation}
p=\left(\begin{array}{cccc}
1&0&0&0\cr 0&1&0&0\cr 0&0&-1&0\cr 0&0&0&-1\cr\end{array}\right).
\label{e16}
\end{equation}
Note that with these choices the operators $\cP$ and $\cT$ commute, $\cP^2={\bf
1}$, and $\cT^2=-{\bf 1}$. Also, the matrices $p$ and $t$ satisfy $[p,t]=0$,
$p^2={\bf 1}$, and $t^2=-{\bf 1}$.

These results are nearly identical with those of Bjorken and Drell \cite{X6} in
their discussion of the operators $\cP$ and $\cT$ for the Dirac equation. In
this text it is shown that when the parity-reflection operator $\cP$ acts on a
four-component spinor, it has the effect of multiplying the spinor by the matrix
$\gamma^0$, which is precisely the matrix $p$ given in (\ref{e16}). Furthermore,
when the time-reversal operator $\cT$ acts on a four-component spinor, it has
the effect of multiplying the spinor by the matrix $i\gamma^1\gamma^3$, which is
the matrix $\tau=it$, with $t$ given in (\ref{e14}). Thus, $\tau^2={\bf 1}$.
It still follows that $\cT^2=-{\bf 1}$ because $\tau$ is imaginary and thus it 
changes sign under complex conjugation.
\vspace{0.1cm}

\subsection{Construction of quadratically nilpotent matrices}
\label{ss32}

Our next task is to construct general classes of quadratically nilpotent
matrices. We know that an $n$-dimensional matrix whose square vanishes
must have a vanishing trace and determinant. Of course, if $n>2$, not all
traceless $n$-dimensional matrices having a vanishing determinant are
quadratically nilpotent. Thus, we propose the following very simple general set
of such matrices: Let the elements in the top row of the matrix be arbitrarily
chosen complex numbers: $a_1,\,a_2,\,a_3,\ldots,a_n$. Next, let the $k$th row
($k>1$) be an arbitrary multiple $b_k$ of the elements in the first row.
This matrix contains $2n-1$ arbitrary complex parameters and by construction its
determinant vanishes.

We then impose the condition that the matrix be traceless:
\begin{equation}
a_1+a_2b_2+a_3b_3+\ldots+a_nb_n=0.
\label{e17}
\end{equation}
The resulting matrix contains $2n-2$ complex parameters and is quadratically
nilpotent. In four dimensions this construction gives the following general
12-parameter complex matrix representation for $\e$:
\begin{equation}
\e=\left(\begin{array}{cccc} -ch-bg-af&f&g&h\cr -a(ch+bg+af)&af&ag&ah\cr
-b(ch+bg+af)&bf&bg&bh\cr -c(ch+bg+af)&cf&cg&ch\cr \end{array}\right).
\label{e18}
\end{equation}

Using the matrix representation in (\ref{e14}) for the time-reversal operator
and the matrix representation $p$ in (\ref{e16}) for the parity-reflection
operator, we obtain from (\ref{e3}) the $\cP\cT$ reflection of $\e$ from the
formula $\eb=-p\,t\,\e^*\,t\,p$ \cite{X7}:
\begin{equation}
\eb=\left(\begin{array}{cccc}
af&a(ch+bg+af)&-ah&ag\cr -f&-ch-bg-af&h&-g\cr -cf&-c(ch+bg+af)&ch&-cg\cr
bf&b(ch+bg+af)&-bh&bg\cr \end{array}\right)^*.
\label{e19}
\end{equation}
One can verify that $\eb^2=0$.

\subsection{Grassmann algebra}
\label{ss33}

Using the $4\times4$ matrix representations for $\e$ in (\ref{e18}) and $\eb$ in
(\ref{e19}), we can now construct the anticommutator $y=\e\eb+\eb\e$. For the
special case in which the parameters $a$, $b$, $c$, $f$, $g$, $h$ are real, the
matrix $y$ has a particularly simple form because the expression $ach-bh+cg+abg+
a^2f+f$ factors out of all 16 matrix elements:
\begin{widetext}
\begin{equation}
y=(ach-bh+cg+abg+a^2f+f)\left(\begin{array}{cccc} -ach-abg-a^2f-f&-ch-bg&h+ag&
ah-g\cr ch+bg&-ach-abg-a^2f-f&ah-g&-h-ag\cr c^2h+bcg+acf-bf&-bch-b^2g-cf-abf&bh
-cg&-ch-bg\cr-bch-b^2g-cf-abf&-c^2h-bcg-acf+bf&ch+bg&bh-cg\cr\end{array}\right).
\label{e20}
\end{equation}
\end{widetext}
Thus, if we choose 
\begin{equation}
f=(bh-ach-cg-abg)/(a^2+1),
\label{e21}
\end{equation}
then all 16 matrix elements of $y$ vanish, and we have found a five-parameter
four-dimensional real matrix representation of $\e$ and $\eb$ for the Grassmann
algebra (\ref{e2}).

There is no choice of parameters for which $y={\bf 1}$. To show that this is
true, we see from (\ref{e20}) that $y_{2,3}=0$ requires that $g=ah$ and that
$y_{2,4}=0$ requires that $h=-ag$. Combining these two equations gives $g(a^2+1)
=0$. Thus, $g=0$ and also $h=0$. It follows that $y_{4,4}=0$ and we conclude
that it is impossible to construct a real matrix representation of $\e$ that
obeys the $\cP\cT$-symmetric fermionic operator algebra (\ref{e1}).

In general, the parameters in (\ref{e18}) and (\ref{e19}) are complex numbers:
$a=a_1+a_2i$, $b=b_1+b_2i$, $c=c_1+c_2i$, $f=f_1+f_2i$, $g=g_1+g_2i$, $h=h_1+
h_2i$. If we set the $(1,2)$ matrix element of the anticommutator matrix $y$ to
$0$, we obtain two equations for the vanishing of the real and imaginary part.
These equations are long and complicated and they are quadratic in all of the
parameters $a_1$, $a_2$, $b_1$, $b_2$, $\ldots$, except for two; surprisingly,
they are {\it linear} in $f_1$ and $f_2$. If we solve this pair of equations
simultaneously, we obtain startlingly simple results for the real and imaginary
parts of $f$:
\begin{eqnarray}
f_1&=&[(a_1c_2-a_2c_1+b_2)h_2+(-a_2c_2-a_1c_1+b_1)h_1\nonumber\\
&&+(-c_2+a_1b_2-a_2b_1)g_2+(-c_1-a_2b_2-a_1b_1)g_1]\nonumber\\
&&/(a_2^2+a_1^2+1),\nonumber\\
f_2&=&-[(a_2c_2+a_1c_1-b_1)h_2+(a_1c_2-a_2c_1+b_2)h_1\nonumber\\
&&+(c_1+a_2b_2+a_1b_1)g_2+(-c_2+a_1b_2-a_2b_1)g_1]\nonumber\\
&&/(a_2^2+a_1^2+1).
\label{e22}
\end{eqnarray}
This is the complex generalization of (\ref{e21}).

Substituting $f_1$ and $f_2$ into $y$, we find that all 16 matrix elements of
$y$ vanish. Thus, we have found a 10-parameter complex Grassmann representation.
An interesting special case is the complex {\it symmetric} representation (there
is no real symmetric representation):
\begin{equation}
\e=\left(\begin{array}{cccc}
-i&-i\alpha&1&\alpha\cr -i\alpha&-i\alpha^2&\alpha&\alpha^2\cr
1&\alpha&i&i\alpha\cr \alpha&\alpha^2&i\alpha&i\alpha^2\cr\end{array}\right),
\label{e23}
\end{equation}
for which $\alpha$ is real. This representation has an obvious quaternionic
structure.
\vspace{.1cm}

\subsection{The Peculiar $\cP\cT$-Symmetric Fermionic case}
\label{ss34}
To construct a fermionic algebra (for which the matrix $y$ is nonvanishing), we
must {\it not} allow $f_1$ and $f_2$ to take the values in (\ref{e22}). The
expressions for the matrix elements of $y$ are extremely complicated, but
(\ref{e22}) indicates a way to proceed. Note that the denominator of (\ref{e22})
is quadratic in $a_1$ and $a_2$. This suggests that we should choose $f_1=0$ and
$f_2=0$ so that we can obtain {\it linear} equations to solve for $a_1$ and
$a_2$. We find that it is simplest to solve simultaneously the real and
imaginary parts of $y_{3,4}=0$ for $a_1$ and $a_2$ and we obtain
\begin{equation}
a_1=N_1/D,\qquad a_2=N_2/D,
\label{e24}
\end{equation}
where
\begin{widetext}
$$N_1=(b_1g_1-b_2g_2+c_1h_1-c_2h_2)(b_1h_1+b_2h_2-c_1g_1-c_2g_2)
+(b_1g_2+b_2g_1+c_1h_2+c_2h_1)(b_1h_2-b_2h_1-c_1g_2+c_2g_1),$$
$$N_2=(b_1h_2-b_2h_1-c_1g_2+c_2g_1)(b_2g_2-b_1g_1-c_1h_1+c_2h_2)
+(b_1h_1+b_2h_2-c_1g_1-c_2g_2)(b_1g_2+b_2g_1+c_1h_2+c_2h_1),$$
$$D=(c_2h_2+b_2g_2-c_1h_1-b_1g_1)^2 +(c_2h_1+b_1g_2+c_1h_2+b_2g_1)^2.$$
\end{widetext}

Amazingly, it is not necessary to solve for any other parameters; we find that 
when we substitute the values of $a_1$ and $a_2$ in (\ref{e24}) into the matrix
$y$ we obtain after massive simplification a stunningly simple expression for
$y$:
\begin{equation}
y=\e\eb+\eb\e=\left(\begin{array}{cccc} -1&0&0&0\cr 0&-1&0&0\cr 0&0&-1&0\cr
0&0&0&-1\cr \end{array}\right).
\label{e25}
\end{equation}
This result is a surprise because $y=-{\bf 1}$ rather than the expected matrix
$\bf 1$. Evidently, it is not possible to achieve the conventional fermionic
algebra in (\ref{e1}) but rather we get the $\cP\cT$-symmetric variant of this
algebra in (\ref{e4}). Indeed, we have found an eight-parameter representation
of this algebra in which the real and imaginary parts of $b$, $c$, $g$, and $h$
are arbitrary. Note that we cannot change the sign in this algebra from $-1$
back to $+1$ by multiplying $\e$ by a complex phase because the time-reversal
operator $\cT$ performs complex conjugation.

We conclude that because the time operator for fermions obeys the equation
$\cT^2=-{\bf 1}$, the fermionic operator algebra for a $\cP\cT$-symmetric system
necessarily picks up an extra minus sign; we must replace the algebra in
(\ref{e1}) by its $\cP\cT$ variant in (\ref{e4}). We interpret the negative sign
in the fermionic algebra (\ref{e4}) as indicating a fundamental change in
character from elliptic to hyperbolic. This is the same interpretation that we
presented at the end of Sec.~\ref{s2} for the case of a $2\times2$ $\cP
\cT$-symmetric matrix representation for $\e$.

CMB is grateful to the Graduate School at the University of Heidelberg for its
hospitality. CMB thanks the U.S.~Department of Energy for financial support.


\begin{thebibliography}{99}
\bibitem{X7} The minus sign in this formula is crucial. It arises because of the
following argument: Let $A$ be a four-dimensional matrix operator and let $v$ be
an arbitrary four-component vector. Then, if we apply $\cT A\cT^{-1}$ to $v$, we
obtain $\cT A\cT^{-1}v=-\cT A\cT v=-\cT(Atv^*)=-t(Atv^*)^*=-tA^*tv$. Since $v$
is arbitrary, we conclude that $\cT A\cT^{-1}=-tA^*t$.
\end{thebibliography}
\end{document}